\documentclass[hyper,letterpaper,12pt]{article}
\usepackage{a4wide}

\usepackage{mathrsfs}
\usepackage{color}
\usepackage{graphicx}
\usepackage{amsmath}
\usepackage{amssymb}
\usepackage{bbm}
\usepackage[
       colorlinks=true,
      filecolor=black,
       anchorcolor=blue,
      linkcolor=blue,
      citecolor=red,
      urlcolor=blue,
       linktocpage=true,
        plainpages=false,
        breaklinks=true,
            pdfstartview=FitH
           ]{hyperref}

\def\I{I }
\def\sc{$\Sigma_c$ }

\def\0{{(0)}}
\def\1{{(1)}}
\def\2{{(2)}}
\def\3O{O(\partial^3)}
\def\gv{{\gamma_v}}
\def\m{\mu}
\def\n{\nu}
\def\l{{\lambda}}
\def\a{{\alpha}}
\def\b{{\beta}}
\def\g{{\gamma}}
\def\de{{\delta}}
\def\o{\omega}
\def\t{\tau}
\def\e{{\epsilon}}
\def\ep{\epsilon}
\def\eps{\epsilon}
\def\D{\partial}
\def\Oo{\Omega}
\def\s{\sigma}
\def\r{\rho}
\def\p{\partial}

\def\D{\partial}
\def\d{\mathrm{d}}
\def\K{\mathcal{K}}

\def\lgb{\mathcal{L}_{GB}}
\def\mathM{\mathcal{M}}

\def\ng{g}

\def\L{Y}

\def\mL{\mathscr{L}}
\def\M{M}
\def\N{N}
\def\vr{\varrho}

\def\vp{\pi}

\def\iu{\!\!\;}

\def\pe{\perp}

\def\erho{\mathrm{e}}
\def\pp{\mathrm{p}}
\def\Pe{\mathrm{P}}
\def\pp{\mathbbm{p}}

\def\Pe{\mathbb{P}}
\def\Ha{\mathbb{H}}

\def\Hij{\mathrm{H}}
\def\bHij{\bar{\mathrm{H}}}
\def\Pij{\mathrm{P}}
\def\erho{\mathbbm{e}}

\def\Beta{\mathbbm{j}}
\def\erho{\mathbbm{e}}

\def\upH{_{^{^{(H)}}}}

\def\bg{\bar g}

\def\bT{\bar T}
\def\bK{\bar K}
\def\bJ{\bar J}
\def\bR{\bar R}
\def\bPe{\bar \Pe}
\def\bPij{\bar \Pij}
\def\bHa{\bar \Ha}
\def\bH{\bar H}
\def\bM{\bar M}
\def\bN{\bar N}
\def\bL{\bar Y}

\def\dT{\delta T}
\def\dK{\delta K}

\def\dM{\delta M}
\def\dN{\delta N}

\def\dvr{\delta\varrho}

\def\dvp{\delta\pi}

\def\ttT{\tilde T}

\def\tg{  g}

\def\tu{  u}
\def\ts{  s}
\def\tT{  T}
\def\tK{  K}
\def\tJ{  J}
\def\tPe{  \Pe}

\def\tHa{  \Ha}
\def\tH{  H}
\def\tM{  M}
\def\tN{  N}
\def\tL{  Y}
\def\tC{  C}
\def\tR{  R}

\def\tn{ n}

\def\ttTGB{{ T}^{_{(GB)}}}
\def\ttPeGB{\mathbb{P}}
\def\ttHaGB{\mathbb{H}}
\def\PijGB{\mathrm{P}}
\def\HijGB{\mathrm{H}}

\def\mbhl{\mathcal {B}(\hell)}

\def\be{\begin{equation}}
\def\ee{\end{equation}}
\def\beq{\begin{eqnarray}}
\def\eeq{\end{eqnarray}}
\def\bea{\begin{eqnarray}}
\def\eea{\end{eqnarray}}
\def\nn{{\nonumber}}
\renewcommand{\(}{\left(}
\renewcommand{\)}{\right)}
\renewcommand{\[}{\left[}
\renewcommand{\]}{\right]}
\newcommand{\bm}[1]{\mbox{\boldmath{$#1$}}}
\newcommand{\bml}[1]{\mbox{\scriptsize\boldmath{$#1$}}}

\def\hu{{\bm u}}
\def\hm{{\bm m}}
\def\hk{{\bm k}}
\def\hell{{\bm{\ell}}}
\def\hn{{\bm n}}

\def\bmell{{\bml{\ell}}}
\def\bmk{{\bml{k}}}
\def\bmn{{\bml{n}}}
\def\bmu{{\bml{u}}}

\usepackage{t1enc}
\def \BE {\begin{equation}}
\def \EE {\end{equation}}
\def \BEAH {\begin{eqnarray*}}
\def \EEAH {\end{eqnarray*}}
\def \BEA {\begin{eqnarray}}
\def \EEA {\end{eqnarray}}
\def \BDM {\begin{displaymath}}
\def \EDM {\end{displaymath}}
\def \BDM {\begin{equation}}
\def \EDM {\end{equation}}

\begin{document}
\title{\bf \Large Petrov type I Condition and Rindler Fluid in Vacuum \\
Einstein-Gauss-Bonnet Gravity}

\author{
Rong-Gen Cai$^{1}$\footnote{E-mail: cairg@itp.ac.cn},
~~Qing Yang$^{1}$\footnote{E-mail: yangqing@itp.ac.cn},
~~Yun-Long Zhang$^{1}$\footnote{E-mail: zhangyl@itp.ac.cn}\\
\\
\small 
$^1$State Key Laboratory of Theoretical Physics, Institute of Theoretical Physics, \\
\small 
Chinese Academy of Sciences, Beijing 100190, People's Republic of China\\
}

\date{\small August 28, 2014}

\maketitle

\begin{abstract}
Recently the Petrov type I condition is introduced to reduce the degrees of freedom in the extrinsic curvature of a timelike hypersurface to the degrees of freedom in the dual Rindler fluid in Einstein gravity. In this paper we show that the Petrov type I condition holds for the solutions of vacuum Einstein-Gauss-Bonnet gravity up to the second order in the relativistic hydrodynamic expansion.
 On the other hand, if imposing the Petrov type I condition and Hamiltonian constraint on a finite cutoff hypersurface, the stress tensor of the relativistic Rindler fluid in vacuum Einstein-Gauss-Bonnet gravity can be recovered with correct first order and second order transport coefficients.
\end{abstract}


\tableofcontents

\section{Introduction}

There has been increasing interest on the holographic duality between fluid dynamics and gravity in the past few years, while the suggestion of such a connection can be dated back to the 1970s proposed by Damour
\cite{Damour1979,Damour:1978cg}. The approach is developed into the membrane paradigm \cite{Price:1986yy}, which relates the black hole evolution and diffusion with those in hydrodynamics~\cite{Kovtun:2003wp,Gourgoulhon:2005ng,Gourgoulhon:2005ch,Eling:2009pb,Eling:2009sj}.
In recent years, along with the progress in the anti-de Sitter/Conformal Field Theory (AdS/CFT) correspondence~\cite{Maldacena:1997re,Gubser:1998bc,Witten:1998qj,Aharony:1999ti},
the dual fluid has been generalized to the conformal fluid living on the boundary of AdS spacetime, which can describe the long wavelength and low frequency limit of conformal field theory~\cite{Policastro:2001yc,Policastro:2002se,Policastro:2002tn}.
In particular, a systematic method to study the duality was proposed in the fluid/gravity correspondence~\cite{Bhattacharyya:2008jc}, which translates problems in fluid dynamics into problems in general relativity. It was then further expanded to arbitrary dimensions in \cite{VanRaamsdonk:2008fp,Haack:2008cp,Bhattacharyya:2008mz} and to non-relativistic hydrodynamics in \cite{Bhattacharyya:2008kq}.

To build up the connection between the fluid/gravity correspondence and membrane paradigm, a timelike hypersurface outside the horizon is introduced to study the universality of the hydrodynamic limit in AdS/CFT correspondence and membrane paradigm~\cite{Iqbal:2008by,Nickel:2010pr,Bredberg:2010ky}. Significantly, the authors in \cite{Bredberg:2010ky} consider the fluid living on the finite cutoff hypersurface from the viewpoint of Wilsonian renormalization, there Dirichlet boundary condition on the hypersurface and the regularity on the horizon are imposed. Then the fluid/gravity correspondence on the cutoff hypersurface
can be generalized to either asymptotically flat \cite{Bredberg:2011jq,Bredberg:2011xw} or de Sitter spacetime \cite{Anninos:2011zn}, and it has been further studied in~\cite{Cai:2011xv,Kuperstein:2011fn,Brattan:2011my,Niu:2011gu,Bai:2012ci,Cai:2012mg,Zou:2013ix,Emparan:2013ila,Zou:2013fua}.
More general discussions in fluid/gravity correspondence can also be found in \cite{Eling:2011ct,Cai:2012vr,Matsuo:2012pi,Kuperstein:2013hqa,Pinzani-Fokeeva:2014cka}, as well as in the frame of  AdS/Ricci-flat correspondence \cite{Caldarelli:2012hy,Caldarelli:2013aaa}.

 In the fluid/gravity duality, one of the most important developments is the so-called Rindler hydrodynamics~\cite{Bredberg:2011jq,Compere:2011dx,Chirco:2011ex,Compere:2012mt,Eling:2012ni,Eling:2012xa,Meyer:2013sva},
where the dual fluid lives on a constant acceleration hypersurface with a flat induced metric.
More interestingly, it is found in~\cite{Lysov:2011xx} that in the near-horizon limit,
instead of the regularity condition on the horizon, imposing the Petrov type \I condition on the hypersurface can reduce the vacuum Einstein equations to the incompressible Navier-Stokes equations in one lower dimensional flat spacetime. It is mathematically much simpler than solving gravitational field equations.
Further study based on this framework can be found in~\cite{Huang:2011he,Huang:2011kj,Zhang:2012uy,Wu:2013kqa,Wu:2013mda,Ling:2013kua}.
From the point of view of degrees of freedom, the Petrov type I condition gives $(p+2)(p-1)/2$ constraints on the
extrinsic curvature of a $p+1$ dimensional timelike hypersurface, or equivalently on 
the dual Brown-York stress tensor.
Then the degrees of freedom of the stress tensor are reduced to be $p+2$, which can be interpreted as energy density, pressure and velocity field of dual fluid~\cite{Lysov:2011xx}. Furthermore the momentum constraint turns out to be the equation of motion of the dual fluid, and the Hamiltonian constraint can be interpreted as the equation of state.

Recently, it has been shown in \cite{Cai:2013uye,Cai:2014ywa} that, the Petrov type \I condition can be used to recover the stress tensor of the dual fluid on the hypersurface order by order under appropriate gauge choice. Without solving the perturbative gravitational field equations, the Rindler fluid in vacuum Einstein gravity can be recovered at least up to the second order in the relativistic hydrodynamic expansion~\cite{Cai:2014ywa}. Note that the stress tensor of Rindler fluid in vacuum Einstein-Gauss-Bonnet gravity is found to be modified by the Gauss-Bonnet coefficient $\a$ in~\cite{Chirco:2011ex,Eling:2012xa}. It is then quite interesting to ask whether the Petrov type I condition holds or not in the vacuum Einstein-Gauss-Bonnet gravity and whether it can be used to recover the dual stress tensor.
In this paper, we find that the Petrov type I condition for the solution of vacuum Einstein-Gauss-Bonnet equations still holds up to the second order in the relativistic hydrodynamic expansion, and that turn the logic around, imposing the Petrov type I condition and Hamiltonian constraint, the stress tensor of the relativistic Rindler fluid can be recovered with correct first order and second order transport coefficients including the Gauss-Bonnet term corrections.
To be specific, in section \ref{RindlerFluid}, we firstly review the Rindler fluid in vacuum Einstein-Gauss-Bonnet gravity, and show that the spacetime with perturbation is at least Petrov type I up to the second order in the relativistic hydrodynamic expansion.
In section \ref{PetrovwithGB}, we give a detailed derivation of the Petrov type I condition on a cutoff hypersurface in vacuum Einstein-Gauss-Bonnet gravity. In section \ref{relativistic}, we turn the logic around and assume the Hamiltonian constraint and Petrov type \I conditiont on a finite cutoff hypersurface to recover the stress tensor of the dual fluid without using the details of the solution. We further study the Petrov type I condition in non-relativistic hydrodynamic expansion in section \ref{Nonrelativistic}, and make the conclusion in section \ref{Conclusion}.

\section{Rindler fluid in Einstein-Gauss-Bonnet gravity}
\label{RindlerFluid}

To study the fluid dual to vacuum Einstein-Gauss-Bonnet gravity,
we begin with the Einstein-Hilbert action on a $(p+2)$ dimensional Lorentz manifold $\mathM$, with the Gauss-Bonnet term $\lgb=R^2-4 R_{\mu\nu} R^{\mu\nu}+R_{\mu\nu\sigma\lambda}R^{\mu\nu\sigma\lambda}$ and appropriate surface term \cite{Myers:1987yn}
\begin{equation}
S=\frac{1}{16\pi G_{p+2}}\int d^{p+2}x \sqrt{-\ng}(R-2\Lambda+\a \lgb)+S_{\p\mathM}.
\end{equation}
where $\a$ is the Gauss-Bonnet coefficient. Varying this action with respect to the metric $\tg_{\mu\nu}$ yields the vacuum Einstein-Gauss-Bonnet field equations,
\begin{align}
&G_{\m\n}+2\a H_{\mu\nu}=0, \qquad G_{\m\n}\equiv R_{\mu\nu}-\frac{1}{2}R g_{\mu\nu},\quad \mu,\nu=0,1,...,p+1,\label{eomGB}\\
&H_{\mu\nu}\equiv RR_{\mu\nu}-2R_{\mu \lambda}R^{\lambda}_{~\nu}-2R^{\sigma\lambda}R_{\mu \sigma \nu \lambda}+R_{\mu}^{~\sigma\lambda\rho}R_{\nu \sigma\lambda\rho}-\frac{1}{4}\ng_{\mu\nu}\lgb.\label{HmnA}
\end{align}
The $p+2$ dimensional Rindler metric
\begin{align}\label{Rindler}
\d s_{p+2}^2=-r \d\t^2+2\d\t \d r+\delta_{ij}\d x^i \d x^j,\quad i,j=1,...,p\, ,
\end{align}
is an exact solution of the field equations \eqref{eomGB}.
On a timelike hypersurface \sc with $r=r_c$, the induced metric is intrinsic flat,
\begin{align}\label{RindlerCut}
\d s_{p+1}^2=\gamma_{ab}\d x^a \d x^b=-r_c \d\t^2+\d x_i \d x^i~,\quad a,b=0,1,...,p\,.
\end{align}
And after setting $16\pi G_{p+2}=1$, the Brown-York stress tensor of Einstein-Gauss-Bonnet gravity on the  cutoff surface \sc can be written as \cite{Davis:2002gn,Cai:2011xv},
\begin{align}\label{STGB}
\ttTGB_{ab}&= -2\(\tK_{ab}-\tK\g_{ab}\)- 4\a\(3\tJ_{ab}-\tJ\gamma_{ab}\),\qquad
\tJ\equiv\gamma^{ab}\tJ_{ab},\\
\tJ_{ab}&\equiv\frac{1}{3}\(2\tK \tK_{ac}\tK^c_{~b}+\tK_{cd}\tK^{cd}\tK_{ab}-2\tK_{ac}\tK^{cd}\tK_{db}-\tK^2\tK_{ab}\).\label{Jab1}
\end{align}
Here $\tK_{ab}$ is the extrinsic curvature of the hypersurface \sc.

\subsection{Rindler fluid in relativistic hydrodynamic expansion}

In order to study the dual fluid on the hypersurface \sc, one introduces the $(p+1)$ independent parameters $u^a=\gv(1,\, v^i)$ and $\pp$, which are slowly varying functions of $x^a=(\t,x^i)$.
Here $\gv$ is fixed through $\gamma_{ab}u^au^b=-1$.
Keep the induced metric on a timelike hypersurface \sc flat and regularity on the future horizon,
the solution of vacuum Einstein-Gauss-Bonnet field equation \eqref{eomGB} up to the second order in the derivative expansion is given by~\cite{Compere:2012mt,Eling:2012ni},
\begin{align}\label{metricGB}
\d \ts^2_{p+2} &=\tg_{\mu\nu} \d x^\mu \d x^\nu= - 2\pp u_a \d x^a \d r+ \tg_{ab} \d x^a \d x^b,\\
\tg_{ab} &= \tg_{ab}^{\0}+\tg_{ab}^{\1}+\tg_{ab}^{\2}+\3O.
\end{align}
The leading order term of $g_{ab}$ in the derivative expansion is
\begin{align}
\tg_{ab}^{\0}&= [1-\pp^2(r-r_c)]u_a u_b+h_{ab},
\end{align}
where the projection tensor $h_{ab}\equiv\gamma_{ab}+u_a u_b$. We can read out the horizon position through $r_h=r_c-1/\pp^2$ with $g^{(0)}_{ab}$ in the case of equilibrium state.
The first order term of $g_{ab}$ in the derivative expansion is
\begin{align}
\tg_{ab}^{\1}&= 2\pp(r-r_c) \[(D \ln\pp) u_a u_b+2 a_{(a} u_{b)}\],
\end{align}
where $D \equiv u^c \p_c$ and the acceleration $a^a\equiv u^b\p_b u^a$.
At the second order in the derivative expansion, the Gauss-Bonnet corrections appear in the metric~\cite{Eling:2012xa},
\begin{align}
u^c u^d \tg^{(2)}_{cd}=& +2(r-r_c)\K_{cd}\K^{cd}+\frac{1}{2}\pp^2 (r-r_c)^2\( \K_{cd}\K^{cd}
+2 a_c a^c \) +\frac{1}{2}\pp^4(r-r_c)^3 \Oo_{cd}\Oo^{cd} \,\nn\\
&+2\a\pp^2(r-r_c) \big(\K_{cd}\K^{cd}-\frac{6}{p}\Oo_{cd}\Oo^{cd}\big)
+ 3\a{\pp^4}(r-r_c)^2\frac{p-2}{p}\Oo_{cd}\Oo^{cd},\\
h_a^c u^d \tg^{(2)}_{cd}=& -2 (r-r_c)h_a^c \p_d \K^d_{\;c}
+\pp^2(r-r_c)^2 \[h_a^b \p_c \K^c_{\; b}-(\K_{ad}+\Omega_{ad})a^d\] \, ,\\
h_a^c h_b^d \tg^{(2)}_{cd}=& +2(r-r_c)\( - \K_a^{~c}\K_{cb}+ 2\K_{c(a}\Omega^c_{~b)}- 2 h_a^c h_b^d D \K_{cd} \)-\pp^2(r-r_c)^2 \Omega_{ac}\Omega^{c}_{~b}\, \nn\\
&+12\a\pp^2(r-r_c)\Big[\Oo_{ac}\Oo^c_{~b}+\frac{1}{p}\(\Oo_{cd}\Oo^{cd}\)h_{ab}\Big]. \label{gab2}
\end{align}
Here the fluid shear and vorticity are defined as
\begin{align}
\K_{ab}\equiv h_a^{c}h_b^{d}\p_{(c} u_{d)},\quad \Oo_{ab}\equiv h_a^{c}h_b^{d}\p_{[c} u_{d]}.
\end{align}
The components of inverse metric up to the second order in the derivative expansion are
\begin{align}
\label{invg2}
\tg^{rr} &=\pp^{-2}\[1+\pp^2(r-r_c) - \(\tg^{(1)}_{cd}+ \tg^{(2)}_{cd} - h^{ab}\tg^{(1)}_{ac} \tg^{(1)}_{bd}\)u^c u^d \] \, ,\nn\\
\tg^{ra} &={\pp}^{-1}\( u^a +h^{ab}\tg^{(1)}_{bc} u^c +h^{ab}\tg^{(2)}_{bc} u^c \) \,, \nn\\
\tg^{ab} &= h^{ab} - h^{ac} h^{bd} \tg^{(2)}_{cd}\, .
\end{align}
one also needs to consider the following constraint equations
\begin{align}\label{constraint}
\p_a u^a &= {2}{\pp}^{-1}\K_{ab}\K^{ab}+O(\p^3),\nn\\
a_a+D_a^\perp\ln \pp &= {2}{\pp}^{-1}h_a^c\p_b \K_c^b+O(\p^3),
\end{align}
with $D_a^{\bot}\equiv h_a^c \p_c$,
so that the metric \eqref{metricGB} solves the vacuum Einstein-Gauss-Bonnet field equations \eqref{eomGB} up to the second order in the derivative expansion.

With the metric \eqref{metricGB} and appropriate gauge choice, the dual stress tensor  $\ttTGB_{ab}$ in the vacuum Einstein-Gauss-Bonnet gravity  on the finite cutoff surface \sc in \eqref{STGB} has been obtained in \cite{Eling:2012ni},
\begin{align}\label{TabGB}
\ttTGB_{ab}=&+\pp h_{ab}-2 \K_{ab}-2 \pp^{-1}\(\K_{ab} \K^{ab}\)u_a u_b\nn\\ &
+\pp^{-1}\[-2\(1+2\a\pp^2\)\K_{ac}\K^c_{~b}-4\K_{c(a}\Oo^c_{~b)}-4\(1+3\a\pp^2\)\Oo_{ac}\Oo^c_{~b}
\right.\nn\\ &\left.\qquad\quad\,-4h_a^c h_b^d\p_c\p_d\ln\pp
-4 \K_{ab}D\ln \pp+ 4 (D_a^\perp\ln \pp)(  D_b^\perp \ln \pp)\].
\end{align}
On the other hand, the general stress tensor $T^{_{(R)}}_{ab}$ for $(p+1)$-dimensional relativistic fluid with vanishing equilibrium energy density is constructed in \cite{Compere:2012mt} as
\begin{align}
T^{_{(R)}}_{ab} =&+\pp h_{ab}-2\eta\K_{ab}+\zeta' (D\ln \pp)u_a u_b
\nn\\&+{\pp}^{-1}\big[ d_1 \K_{cd}\K^{cd} + d_2 \Omega_{cd}\Omega^{cd} + d_3(D \ln \pp )^2
+d_4 DD \ln \pp +d_5 (D_{\bot} \ln \pp )^2\big] u_a u_b
\nn\\&
+{\pp}^{-1}\big[ c_1 \K_{ac}\K^c_{~b}
+ c_2 \K_{c(a}\Omega^c_{~b)}
+ c_3 \Omega_{ac}\Omega^c_{~b}
+c_4 h_a^c h_b^d\D_c\D_d \ln \pp
+ c_5 \K_{ab}\,D\ln \pp\nn\\&\qquad~~~
 + c_6 D^\perp_a \ln \pp \,D^\perp_b\ln \pp \big]  \, .\label{TRab}
\end{align}
Compare $\ttTGB_{ab}$ in \eqref{TabGB} with $T^{_{(R)}}_{ab}$, one can read out the holographic transport coefficients of Rindler fluid in vacuum Einstein-Gauss-Bonnet gravity as
\begin{align}
\zeta' =&\, 0 ,\qquad \eta = 1,\qquad d_1 = -2,\qquad d_2=d_3=d_4=d_5 = 0, \nn\\
c_1=&-2(1+2\alpha\pp^2),\quad c_3=-4(1+3\alpha\pp^2),\quad c_2 = c_4 = c_5 = -c_6 = -4\,.\label{coefficientsGB}
\end{align}
It turns out that there are no Gauss-Bonne corrections to the shear viscosity $\eta$ and the parameter $\zeta'$, the latter measures variations of the energy density. The Gauss-Bonnet corrections appear in the second order transport coefficients $c_1$ and $c_3$.

\subsection{The solution is Petrov type \I}

The Petrov type classification of Weyl tensor in higher dimensions is summarized in Appendix \ref{APetrovtype}.
In this subsection, we will show that the Weyl tensors $\tC_{\mu\nu\alpha\beta}$ of the metric $g_{\mu\nu}$ in \eqref{metricGB} is at least Petrov type \I.

Choose $(p+2)$ Newman-Penrose-like vector fields, which include two null vectors $\hell^2=\hk^2=0$, and $p$ orthonormal space-like vectors $\hm_i$. The null vectors obey $\hell_\mu\hk^\mu=1$ and all other products with  $\hm_i$($i=1,...p$) vanish. Define
\begin{align}\label{Pijr}
\tPe^{(r)}_{ij}\equiv 2\tC_{(\bmell)i(\bmell)j}\equiv2\hell^\mu \hm_i^{\;\nu}
\hell^\alpha \hm_j^{\;\beta}\tC_{\mu\nu\alpha\beta}.
\end{align}
Then the Weyl tensor $\tC_{\mu\nu\alpha\beta}$ is at least Petrov type \I if there exists a frame $\{\hell, \hk, \hm_i\}$ such that $\tPe^{(r)}_{ij}=0$. A special kind of frame has been chosen in \cite{Cai:2014ywa}.
If we denote $\hn$ as the spacelike unit normal vector of a constant $r$ hypersurface, $\hu$ is the normalized $(p+2)$ velocity along with the hypersurface, the two null vector fields can be chosen as their combinations
\begin{align} \label{frame}
\sqrt{2}\hell=-\hn+ \hu,\quad \sqrt{2}\hk=-\hn - \hu.
\end{align}
For the remaining orthonormal spatial vectors $\hm_i$, there exists still a freedom to choose.
Consider the fact that $m_i^{\;a}m^i_{\;b}=h^a_b =\delta^a_b+u^a u_b$, and
\begin{align}
m_i^{\;a} &= \de_i^{\;a}+r_c^{-1/2}u_i\de_\t^a+(1+r_c^{1/2}\gv)^{-1}u_i u^j\de_j^a,\nn\\
m^i_{\;a} &= \de^i_{\;a}-r_c^{+1/2}u^i\de^\t_a+(1+r_c^{1/2}\gv)^{-1}u^i u_j\de^j_a,
\end{align}
the components of the frame have been chosen as follows~\cite{Cai:2014ywa},
\begin{align}\label{mi}
\sqrt{2}\hell^\mu&=-\hn^r\delta^\mu_r-(\hn^a-\hu^a)\delta^\mu_a=(\tg^{rr})^{1/2}\delta^\mu_r,\nn\\
\sqrt{2}\hk^\mu&=-\hn^r\delta^\mu_r-(\hn^a+\hu^a)\delta^\mu_a=-(\tg^{rr})^{1/2}\(\delta^\mu_r+2\tg^{ra}\delta^\mu_a\),\nn\\
\hm_i^{~\mu}&=\hm_i^{~a}\delta_a^\mu=\big(m_i^{~a}-\frac{1}{2} m_i^{~b} g^{\2}_{bc}h^{ca}\big)\delta_a^\mu.
\end{align}
And the components with subscript  index are
\begin{align}\label{downindex}
\sqrt{2}\hell_\m=&-(\hn_r-\hu_r)\delta_\m^r+\hu_a\delta^a_\mu=(\tg^{rr})^{-1/2}\pp u_a\delta^a_\mu,\nn\\
\sqrt{2}\hk_\m=&-(\hn_r+\hu_r)\delta_\m^r-\hu_a\delta^a_\mu=-2(\tg^{rr})^{-1/2}\delta_\m^r-(\tg^{rr})^{-1/2}\pp u_a\delta^a_\mu,\nn\\
\hm_{\mu}^{~i}=&\[m_{a}^{~i}+u_{a}u^b(\tg^{(1)}_{bc}+\tg^{(2)}_{bc})h^{cd}m_{d}^{~i}-\frac{1}{2}h_{a}^{b}\tg^{(2)}_{bc}h^{cd} m_{d}^{~i}\]\delta^a_\mu.
\end{align}
Up to order $\p^2$, one can check that $\tg_{\mu\nu}\hm^\mu_i\hm^\nu_j=\delta_{ij}$
is satisfied, and  metric (\ref{metricGB}) as well as its inverse (\ref{invg2}) can be decomposed as
\begin{align}\label{gframe}
\tg_{\mu\nu}=2 \hell_{(\mu} \hk_{\nu)} +\de_{ij}\hm^i_{\;\mu}\hm^j_{\;\nu},\qquad
\tg^{\mu\nu}=2 \hell^{(\mu} \hk^{\nu)} +\de^{ij}\hm_i^{\;\mu}\hm_j^{\;\nu}.
\end{align}

To check the Petrov type \I condition $\tPe^{(r)}_{ij}=0$ of the Weyl tensor,
we introduce another covariant formula $\tPe^{{(r)}}_{ab}$, which is defined as
\begin{align}\label{Clilj}
\tPe^{{(r)}}_{ab}\equiv 2 h_a^c h_b^d\tC_{(\bmell)c(\bmell)d}= \hn^r h_a^c\hn^r  h_b^d C_{rcrd},\quad
\tPe^{(r)}_{ij}= \hm_i^{\;a} \hm_j^{\;b}\tPe^{(r)}_{ab}.
\end{align}
Then after a straightforward calculation of the Weyl tensors with metric (\ref{metricGB}), we find
\begin{align}\label{Rrarb}
\tPe^{{(r)}}_{ab}=&-\tg^{rr}\(\frac{1}{2}h_a^c h_b^d \partial_r^2 \tg^{(2)}_{cd}+\pp^2\Oo_{ac}\Oo^c_{\;b}\)+\3O.
\end{align}
Considering $\tg^{(2)}_{cd}$ with Gauss-Bonnet corrections in (\ref{gab2}), we can conclude that $\tPe^{{(r)}}_{ab}=O(\p^3)$ at arbitrary $r$, which also indicates $\tPe^{(r)}_{ij}=O(\p^3)$ at every spacetime point in  \eqref{metricGB}.
As a result, we have shown that the Weyl tensor or the spacetime with metric \eqref{metricGB} is at least Petrov type \I  up to $\p^2$, even when the Gauss-Bonnet term is included.

\section{Petrov type \I condition on the hypersurface $\Sigma_c$ }
\label{PetrovwithGB}

The Petrov type \I condition is introduced to reduce the degrees of freedom in the extrinsic curvature of the hypersurface \sc to the degrees of freedom in the dual fluid on \sc in \cite{Lysov:2011xx}.
On this hypersurface, the covariant Petrov type \I condition is defined as \cite{Cai:2014ywa},
\begin{align}\tPe_{ab}\equiv \tPe^{(r_c)}_{ab}=2 h_a^c h_b^d\tC_{(\bmell)c(\bmell)d}|_{\Sigma_c}=0.\label{PetrovCut}
\end{align}
With \eqref{frame} and consider the fact that
\begin{align} \label{Weylell}
2\tC_{(\iu \bmell\iu)c(\iu \bmell\iu)d}=\tC_{(\iu \bmu\iu)c(\iu \bmu\iu)d}-\tC_{(\iu \bmu\iu)c(\iu \bmn\iu)d}
-\tC_{(\iu \bmu\iu)d(\iu \bmn\iu)c}+\tC_{(\iu \bmn\iu)c(\iu \bmn\iu)d},
 \end{align}
we need to rewrite the Weyl tensor in terms of the extrinsic curvature $K_{ab}$,
through using the Gauss-Codazzi equations on the intrinsic flat hypersurface \sc.
Thus, we firstly define the following notations
\begin{align}\label{Mabcd}
\tM_{abcd}\equiv\g_a^\a \g_b^\b\g_c^\g \g_d^\delta \tR_{\a\b\g\de}&=\tK_{ad}\tK_{bc} -\tK_{ac}\tK_{bd},\nn\\
\tN_{abc}\equiv\g_a^\a \g_b^\b \g_c^\g \tn^\de \tR_{\a\b\g\de} &= \p_a\tK_{bc}-\p_b\tK_{ac},  \nn\\
\tL_{ab}\equiv\g_a^\a \tn^\b \g_b^\g \tn^\de \tR_{\a\b\g\de}&=\tK \tK_{ab}-\tK_{ac}\tK^c_{~b}+\g_a^\a \g_b^\g \tR_{\a\g},
\end{align}
with $\gamma^\alpha_a=\delta^\alpha_a-n_a n^\alpha=\delta^\alpha_a$,
as well as their  contractions,
\begin{align}\label{MNY}
\tM_{ac}&\equiv\g^{bd}\tM_{abcd}= \tK_{ab}\tK^b_{~c}-\tK \tK_{ac},\,\quad
\tN_{b}\equiv\g^{ac}\tN_{abc}= \p_a\(\tK_{~b}^a-\tK\g_{~b}^a\),\nn\\
\tM&\equiv\g^{ac}\tM_{ac}= \tK_{ab}\tK^{ab}-\tK^2,\qquad\quad~
\tL\equiv\g^{ac}\tL_{ac}=-\tM+\g^{\a\b} \tR_{\a\b}.
\end{align}
Then using the equations of motion (\ref{eomGB}) which lead to
\begin{align}\label{GBeom1}
\tR_{\mu\nu}=-\frac{2}{p}\a \tH \tg_{\mu\nu}-2\a \tH_{\mu\nu},
\quad \tR=\frac{4}{p}\a \tH, \quad
H\equiv H_{\mu\nu}g^{\m\n},
\end{align}
we can obtain the projections of the Weyl tensor on the hypersurface \sc,
\begin{align}\label{typeI}
 \g_a^\a \g_b^\b\g_c^\g \g_d^\delta \tC_{\a\b\g\de}&= \tM_{abcd}-\frac{8\a \tH}{p(p+1)}\g_{a[c}\g_{d]b}
 +\a\frac{4}{p}\g_{a}^{\a}\g_{b}^{\b}\g_{c}^{\g}\g_{d}^{\delta}(g_{\a[\g}\tH_{\delta]\b}-g_{\b[\g}\tH_{\delta]\a}),\nn\\
\g_a^\a \g_b^\b \g_c^\g n^\de \tC_{\a\b\g\de} &= \tN_{abc}+\a\frac{4}{p}\g_{a}^{\a}\g_{b}^{\b}\g_{c}^{\g}n^{\delta}(g_{\a[\g}\tH_{\delta]\b}-g_{\b[\g}\tH_{\delta]\a}), \nn\\
\g_a^\a n^\b \g_c^\g n^\de \tC_{\a\b\g\de}&=\tL_{ac}-\frac{4\a \tH}{p(p+1)}\g_{ac}+\a\frac{4}{p}\g_{a}^{\a}n^\b\g_{c}^{\g}n^{\delta}\(g_{\a[\g}\tH_{\delta]\b}-g_{\b[\g}\tH_{\delta]\a}\)\,.
\end{align}
This is similar to the derivation in \cite{Zhang:2012uy} for the case of Einstein gravity with matter.
Then put \eqref{typeI} into \eqref{PetrovCut} and consider \eqref{Weylell},
we obtain $\ttPeGB_{ab}=\tPe^{(\a)}_{ab}+\de\tPe^{_{(H)}}_{ab}$, where
\begin{align}
\tPe^{(\a)}_{ab}&\equiv
\tM^{\pe}_{(\iu u\iu)a(\iu u\iu)b}+2\tN^{\pe}_{(\iu u\iu)(ab)}-\tM^{\pe}_{ab},\label{Paba}\\
\de\tPe^{_{(H)}}_{ab}&\equiv-2\a \tH^{\pe}_{ab} +2\a p^{-1}\[\tH_{(\iu n\iu)(\iu n\iu)}-2\tH_{(\iu n\iu)(\iu u\iu)}+\tH_{(\iu u\iu)(\iu u\iu)}+\tH\]h_{ab}.\label{PabH}
\end{align}
For convenience, we here have defined
\begin{align}\label{MNYbot}
\tM^{\pe}_{(\iu u\iu)a(\iu u\iu)b}&=h_a^m h_b^n \tM_{c m d n} u^c u^d,\quad
\tN^{\pe}_{(\iu \tu\iu)(ab)}=h_{(a}^m h_{b)}^n\tN_{ c mn} u^c,\quad
\tM^{\pe}_{ab}=h_a^m h_b^n\tM_{mn},
\end{align}
as well as
\begin{align}
\tH^{\pe}_{ab}&\equiv \tH_{\m\n}  \g^\m_c \g^\n_d h_a^c h_b^d,\quad
\tH_{(\iu n\iu)(\iu n\iu)}\equiv \tH_{\m\n} n^\m n^\n,\nn\\
\tH_{(\iu u\iu)(\iu u\iu)}&\equiv \tH_{\m\n}  \g^\m_a \g^\n_b u^a u^b,\quad\,
\tH_{(\iu n\iu)(\iu u\iu)}\equiv \tH_{\m\n} n^\m \g^\n_b u^b.
\end{align}

On the other hand, the Hamiltonian constraint for the vacuum Einstein-Gauss-Bonnet field equations \eqref{eomGB} is
\begin{align}\label{Hamiltonian1}
\Ha\equiv&-2(G_{\mu\nu}+2\a H_{\m\n})n^\mu n^\nu=0.
\end{align}
With the decomposition of the Riemann tensor in Appendix \ref{Projections},
we obtain $\Ha=\Ha^{(\a)}+\de\Ha^{\upH}$, where~\cite{Maeda:2003vq}
\begin{align}
\tHa^{(\a)}\equiv \M,\quad
\de\Ha^{\upH}\equiv \a\(\M^2-4M_{ab}\M^{ab}+\M_{abcd}\M^{abcd}\).\label{Hama}
\end{align}
While the momentum constraint for the equations of motion \eqref{eomGB} turns out to be
\begin{align}
\p^a\ttTGB_{ab}\equiv&-2(E_{\mu\nu}+2\a H_{\m\n})n^\mu \g^\nu_b=0,
\end{align}
where $\ttTGB_{ab}$ is the one given in \eqref{STGB}.

Notice that $\tPe^{(\a)}_{ab}$ in \eqref{Paba} has become the hypersurface function of extrinsic curvature $K_{ab}$,  but it is not true for $\de\tPe^{_{(H)}}_{ab}$ in \eqref{PabH}. For example,  we can see from~\cite{Maeda:2003vq} that the term
\begin{align}\label{YabMab}
\tL_{ab}=-\M_{ab}+\g_a^\m \g_b^\n R_{\m\n}=-\mL_n K_{ab}+K_{ac}K^{c}_{~b}
\end{align}
appears in $2\a \tH^{\pe}_{ab}$,
 $Y_{ab}$ can not be obtained only from the  extrinsic curvature $K_{ab}$ and other intrinsic quantities, because additional information of the bulk gravity such as $R_{\mu\nu}$, or the analytic continuation of $K_{ab}$ out of the hypersurface along $n$ is needed.
Thus the  purpose of Petrov type \I condition that gives constraints to the extrinsic curvature  can not be realized in this scene.
However, if we consider only the small Gauss-Bonnet parameter $\a$ limit,
and take the Petrov type \I condition up to the first order in the $\a$ expansion,
the above difficulty can be relieved.

To see this, we firstly define all the quantities with bars have the same formulas as those without bars when $\a=0$.
Then put \eqref{GBeom1} into \eqref{YabMab} and \eqref{HmnA}, we obtain $\bL_{ab}=-\bM_{ab}$,
as well as
\begin{align}\label{bHmn}
\tH_{\mu\nu}=\bH_{\m\n}+O(\a),\quad
\bH_{\m\n}\equiv\bR_{\mu}^{~\sigma\lambda\rho}\bR_{\nu \sigma\lambda\rho}-\frac{1}{4}\(\bR^{\kappa\sigma\lambda\rho}\bR_{\kappa \sigma\lambda\rho}\)\bg_{\m\n}.
\end{align}
With the calculations in Appendix \ref{Projections}, the equation \eqref{PabH} becomes $\de\tPe^{_{(H)}}_{ab}=\de\bPe^{_{(H)}}_{ab}+O(\a^2)$, where
$\de\bPe^{_{(H)}}_{ab}$ is the first order in the small $\a$ expansion that
\begin{align}
\de\bPe^{_{(H)}}_{ab}&\equiv- 2\a \bH^{\bot}_{ab} + 2\a p^{-1}h_{ab}\[\bH_{(\iu n\iu)(\iu n\iu)}-2\bH_{(\iu n\iu)(\iu u\iu)}+\bH_{(\iu u\iu)(\iu u\iu)}+\bH\]\\
&=-2\a h_a^mh_b^n\(\bM_{m}^{~cde}\bM_{n cde}+2\bN_{m}^{~cd}\bN_{n cd}+\bN^{cd}_{~~m}\bN_{cd n}+2\bM_{m}^{~d}\bM_{n d}\) \nn\\
&~~~+\a p^{-1} h_{a b}\Big[2\(\bM_{(\iu u\iu)}^{~~cde}\bM_{(\iu u\iu) cde}+2\bN_{(\iu u\iu)}^{~cd}\bN_{(\iu u\iu) cd}+\bN^{cd}_{~~(\iu u\iu)}\bN_{cd (\iu u\iu)}+2\bM_{(\iu u\iu)}^{~~d}\bM_{(\iu u\iu) d}\)\nn\\
&~~~+4(\bM_{(\iu u\iu)cde}\bN^{dec}-2 \bM^{cd} \bN_{(\iu u\iu)cd}) +\(\bM^{cdef}\bM_{cdef}+6\bN^{cde} \bN_{cde} +8\bM^{cd}\bM_{cd}\)\Big].\label{tPabH}
\end{align}
Now we can say that $\de\bPe^{_{(H)}}_{ab}$ is a  function of $K_{ab}$, $\gamma_{ab}$ as well as $u_a$.
On the other hand, notice that the extrinsic curvature $\tK_{ab}$ can be decomposed as
\begin{align}\label{TKbK}
\tK_{ab}=\bK_{ab}+\dK^{(\a)}_{ab}+O(\a^2),
\end{align}
where $\bK_{ab}$ is the contribution from vacuum Einstein gravity, and $\dK^{(\a)}_{ab}$ includes the terms from the Gauss-Bonnet term  at first order in small $\a$ expansion.
Then from \eqref{Paba} we have $\tPe^{(\a)}_{ab}=\bPe_{ab}+\de\tPe^{(\a)}_{ab}+O(\a^2)$, where
\begin{align}
\bPe_{ab}&\equiv
\bM^{\pe}_{(\iu u\iu)a(\iu u\iu)b}+2\bN^{\pe}_{(\iu u\iu)(ab)}-\bM^{\pe}_{ab},\label{bPab0}\\
\de\tPe^{(\a)}_{ab}&=\dM^{\pe (\a)}_{(\iu u\iu)a(\iu u\iu)b}+2\dN^{\pe (\a)}_{(\iu u\iu)(ab)}-\dM^{\pe(\a)}_{ab}.
\end{align}
Finally, the covariant Petrov type \I condition \eqref{PetrovCut} up to the first order in small $\a$ becomes
\begin{align}\label{ttPeGB}
\ttPeGB_{ab}\equiv\bPe_{ab}+\de\tPe^{(\a)}_{ab}+\de\bPe^{_{(H)}}_{ab}=0.
\end{align}
Similarly,  the  Hamiltonian constraint \eqref{Hamiltonian1} up to the first order in small $\a$ becomes,
\begin{align}\ttHaGB=\bHa+\de\tHa^{(\a)}+\de\bHa^{\upH}=0,\label{ttHaGB}
\end{align}
where
\begin{align}
\bHa&\equiv \bM,\qquad
\de\Ha^{(\a)}\equiv\de\M^{(\a)},\\
\de\bHa^{\upH}\!\!&\equiv \a\(\bM^2-4\bM_{ab}\bM^{ab}+\bM_{abcd}\bM^{abcd}\).\label{tHaH}
\end{align}

With the expansion of $K_{ab}$ in \eqref{TKbK},
the Brown-York stress tensor \eqref{STGB}
can also be expanded as
\begin{align}\label{BYcut}
\ttTGB_{ab}&\equiv\bT_{ab}+\de\tT_{ab}+O(\a^2),\\
\bT_{ab}&\equiv -2(\bK_{ab}-\bK\g_{ab}),\qquad \de\tT_{ab}= \de\tT^{(\a)}_{ab}+\de\bT^{_{(J)}}_{ab}, \qquad ~~
\end{align}
where $\bT_{ab}$ is just the Brown-York stress tensor of Einstein gravity, and $\de\tT_{ab}$ comes from the Gauss-Bonnet term at the first order in small $\a$,
\begin{align}
\de\tT^{(\a)}_{ab}\equiv-2\(\dK^{(\a)}_{ab}-\dK^{(\a)}\g_{ab}\),\quad
\de\bT^{_{(J)}}_{ab}\equiv-4\a (3\bJ_{ab}-\bJ\gamma_{ab}).\label{dTabJ}
\end{align}
In the following section, with the Petrov type \I condition \eqref{ttPeGB} and Hamiltonian constraint \eqref{ttHaGB}, as well as the stress tensor \eqref{BYcut}, we will directly recover the stress tensor \eqref{TabGB} of Rindler fluid in vacuum Einstein-Gauss-Bonnet gravity.

Notice that in the Einstein gravity, $\bK_{ab}$ can be expressed in terms of its Brown-York stress tensor through $\bT_{ab}=2(\bK\gamma_{ab}-\bK_{ab})$. But if we consider the Gauss-Bonnet corrections in \eqref{STGB}, as the cube terms of $K_{ab}$ appear in $J_{ab}$,
 one cannot obtain the extrinsic curvature $K_{ab}$ in terms of the stress tensor $\ttTGB_{ab}$ in (\ref{BYcut}) at finite $\a$. But, up to the first order in small $\a$ and from \eqref{BYcut}, we can have
\begin{align}\label{KabTab}
2\bK_{ab}=&-\bT_{ab}+p^{-1}\bT\g_{ab},\\
2\dK^{(\a)}_{ab}=&-\dT_{ab}+p^{-1}\dT\g_{ab}-4\a\(3\bJ_{ab}-2p^{-1}\bJ\g_{ab}\),\label{dKabTab}
\end{align}
such that the  Petrov type \I condition on the hypersurface can also be expressed in terms of the Brown-York stress tensor in Einstein-Gauss-Bonnet gravity $\ttTGB_{ab}=\bT_{ab}+\de\tT_{ab}$.
Although it is not necessary in our next section \ref{SecPetrovGB},
the formulas in terms of the stress tensor would be much more in accord with the original purpose of the Petrov type I condition \cite{Lysov:2011xx}.
This also gives us the other motivation to take the small $\a$ limit, and we will use this strategy when study the Petrov type \I condition in the non-relativistic hydrodynamic expansion in section \ref{Nonrelativistic}.

\section{From Petrov type \I condition to Rindler fluid}
\label{relativistic}

In this section, we will show how to recover the stress tensor dual to the bulk metric in (\ref{metricGB}) by use of the Petrov type \I
condition without the details of the solution (\ref{metricGB}). We firstly set $\a=0$ to obtain the Rindler fluid in vacuum Einstein gravity from Prtrov type \I condition and Hamiltonian constraint. Then regarding $\a$ as a small parameter, the Gauss-Bonnet corrections to the stress tensor up to first order in small $\a$ can also be obtained naturally.

\subsection{Recover the Rindler fluid in vacuum Einstein gravity}

Firstly, setting $\a=0$ in \eqref{ttPeGB}, we have the Petrov type \I condition on the finite cutoff hypersurface \sc in the vacuum Einstein gravity,
\begin{align}\label{bPabI}
\bPe_{ab}&\equiv
\bM^{\pe}_{(\iu \tu\iu)a(\iu \tu\iu)b}+2\bN^{\pe}_{(\iu \tu\iu)(ab)}-\bM^{\pe}_{ab}=0,
\end{align}
where similar to \eqref{MNYbot}, we have defined
\begin{align}\label{bMNYbot}
\bM^{\pe}_{(\iu \tu\iu)a(\iu \tu\iu)b}&
=h_a^m h_b^n\(\bK_{cm}\bK_{dn}- \bK_{cd}\bK_{mn}\) u^c u^d ,\nn\\
\bN^{\pe}_{(\iu \tu\iu)(ab)}&
=h_{(a}^m h_{b)}^n\(u^c \p_c \bK_{mn} -u^c \p_{m} \bK_{n c}\),\nn\\
\bM^{\pe}_{ab}&
=-h_a^m h_b^n \(\bK \bK_{mn}-\bK_{mc}\bK_{~n}^c\).
\end{align}
On the other hand, from \eqref{KabTab}, we have
\begin{align}\label{TtoK}
2\bK_{ab}=-\bT_{ab}+p^{-1}{\bT}\gamma_{ab},\qquad 2\bK=p^{-1}{\bT}.
\end{align}
Then we can reach the covariant Petrov type \I condition that~\cite{Cai:2014ywa}
\begin{align}\label{petrovP}
4\bPe_{ab}&=h_a^m h_b^n\big[\({\bT}_{mc}{\bT}_{nd}- {\bT}_{mn}{\bT}_{cd}\)u^c u^d
- {\bT}_{mc}\bT^c_{~n}-4 u^c \partial_{c}{\bT}_{mn}
+ 4 u^c\partial_{(m}\bT_{n)c}\big]\nn\\
&\quad+{p^{-2}}\big[\bT({\bT}+p \,{\bT}_{cd}u^c u^d)+ 4 p\, u^c \partial_{c}{\bT}\big]h_{ab}=0.
\end{align}

Now we decompose the arbitrary stress tensor $\bT_{ab}$ associated with a $(p+1)$-velocity $u_a$ as
\begin{align}
\bT_{ab}=\erho u_au_b+2\Beta_{(a}u_{b)}+\Pi_{ab},\quad \bT= -\erho+\Pi.\label{arbST}
\end{align}
where we have defined
\begin{align}
\erho\equiv \, \bT_{ab} u^au^b,\quad \Beta_{a}\equiv -h_a^c\bT_{cd}u^d,\quad
\Pi_{ab}\equiv \, h_a^ch_b^d\bT_{cd},\quad\Pi\equiv  \Pi_{ab}h^{ab}.
\end{align}
Substituting  (\ref{arbST}) into (\ref{petrovP}) we have
\begin{align}\label{PetrovEc}
4 \bPe_{ab}&\equiv-\erho{\Pi}_{\,ab}+2\Beta_a\Beta_b-{\Pi}_{\,ac}{\Pi^c}_{b}
-8 a_{(a}\Beta_{b)}-{4 }h_a^c h_b^d D {\Pi}_{\,cd}
-4\erho\K_{ab}-4D^{\bot}_{(a}\Beta_{b)}
 - 4 \Pi_{(a}^{~~c}D^{\bot}_{b)} u_{c} \nonumber\\
&\quad\,+{p^{-2}}\left[\Pi^2+(p-2)\erho \Pi-(p-1)\erho^2+ 4 p\,D {(\Pi-\erho)}\right]{h}_{\,ab}=0.
\end{align}
Similarly, when $\a=0$, the Hamiltonian constraint in \eqref{ttHaGB} becomes
\begin{align}\label{HamiltonianEc}
4\bHa &\equiv p\bT_{ab}\bT^{ab}-\bT^2=\,2\erho \Pi+(p-1)\erho^2-2p \Beta_a\Beta_b h^{ab}+p\Pi_{ab}\Pi^{ab}-\Pi^2=0.
\end{align}
Expanding the undetermined stress tensor $\bT_{ab}$ in \eqref{arbST} in terms of the derivative expansion parameter $\p$ as
\begin{align}
   \erho    &= \erho^{(0)}+\erho^{(1)}+\erho^{(2)}+O(\p^3),\nonumber\\
 \Beta_a    &=  \Beta_a ^{(0)}+ \Beta_a ^{(1)}+ \Beta_a ^{(2)}+O(\p^3),\nonumber\\
\Pi_{ab}    &=  \Pi_{ab}  ^{(0)}+ \Pi_{ab}  ^{(1)}+\Pi_{ab}   ^{(2)}+O(\p^3),\nonumber\\
    \Pi\, &= \Pi^{(0)}+  \Pi^{(1)} + \Pi^{(2)} +O(\p^3),\label{Tabepsilon}
\end{align}and assuming that the zeroth order of the stress tensor has the same form
as that in the Rindler fluid \eqref{TRab},
\begin{align}\label{zeroth}
\erho^{(0)}=0,\qquad \Beta_a ^{(0)}=0,\qquad  \Pi_{ab}  ^{(0)}=\pp h_{ab},\qquad \Pi^{(0)}=p\pp,
\end{align}
we can recover the first and second order terms of total stress tensor \eqref{TabGB} with $\a=0$,
by imposing the Hamiltonian constraint \eqref{HamiltonianEc} and Petrov type \I condition \eqref{PetrovEc}.
As there is an arbitrary for frame choice of the fluid velocity, we define the relativistic fluid velocity $u^a$ such that $\Beta_a=u^c\bT_{cd}h_{a}^d\equiv 0$ at arbitrary orders, and choose appropriate isotropy gauge that there is no higher order correction to the term which is proportional to $h_{ab}$, that is only $\pp h_{ab}$ appears in the stress tensor \cite{Compere:2012mt}.
To be specific, we can go as follows.

{\bf i) First order.}

We put \eqref{Tabepsilon} and \eqref{zeroth} into the Hamiltonian constraint \eqref{HamiltonianEc} and Petrov type \I condition \eqref{PetrovEc}, and then expand them in the derivative expansion. Assuming $\Beta_a ^{(1)}=0$, at the first order, we have
\begin{align}
\bHa^{(1)}=0&\Rightarrow \erho^{(1)}=0,\\
\bPe_{ab}^{(1)}=0&\Rightarrow \Pi_{ab}^{(1)}=-2\K_{ab}+p^{-1}\(\Pi^{(1)}-\erho^{(1)}\)h_{ab}.\label{Piab1}
\end{align}
Choosing the isotropy gauge such that $\Pi^{(1)}=\erho^{(1)}=0$, we reach $\Pi_{ab}^{(1)}=-2\K_{ab}$.

{\bf ii) Second order.}

With the results in the first order and assuming $\Beta_a ^{(2)}=0$, we can obtain the second order terms
through
\begin{align}
\bHa^{(2)}=0&\Rightarrow \erho^{(2)}=-2 \pp^{-1} \K_{ab} \K^{ab},\\
\bPe_{ab}^{(2)}=0&\Rightarrow \Pi_{ab}^{(2)}
=\pp^{-1}\big[2\K_{ac}\K^c_{~b}-4\K_{c(a}\Oo^c_{~b)}+4h_a^c h_b^d D\K_{cd}\big]+p^{-1}\(\Pi^{(2)}-\erho^{(2)}\)h_{ab}.
\end{align}
Choosing the isotropy gauge such that $\Pi^{(2)}=\erho^{(2)}=-2\pp^{-1}\K_{ab}\K^{ab}$,
and employing the derivatives of momentum constraint equation \eqref{constraint} which lead to the identities,
\begin{align}
h^c_a h^d_b D\mathcal K_{cd} &= - h_{a}^c h_{b}^d \p_c \p_d \ln \pp - \mathcal K_{ab}  D \ln \pp + D^\perp_a \ln \pp D^\perp_b \ln \pp
-\mathcal K_a^{\; c}\mathcal K_{cb}  -\Omega_a^{\; \, c}\Omega_{cb} \nonumber
+O(\p^3),\label{rel4}\\
 h^{cd}D\mathcal K_{cd} &= D\mathcal K = O(\p^3),
\end{align}
%
we finally reach the stress tensor up to the second order in the derivative expansion,
\begin{align}
\bT_{ab}&=+\pp h_{ab}+\(\erho^{\1}+\erho^{\2}\) u_a u_b+\Pi_{ab}^{(1)}+\Pi_{ab}^{(2)}\label{bTab2}\\
&=+\pp h_{ab}-2 \K_{ab}-2 \pp^{-1}\(\K_{ab} \K^{ab}\)u_a u_b
+\pp^{-1}\[-2\K_{ac}\K^c_{~b}-4\K_{c(a}\Oo^c_{~b)}\right.\nn\\ 
&\quad\left.-4\Oo_{ac}\Oo^c_{~b}-4h_a^c h_b^d\p_c\p_d\ln\pp
-4 \K_{ab}D\ln \pp+ 4 (D_a^\perp\ln \pp)(  D_b^\perp \ln \pp)\].\label{Tcovab2}
\end{align}
Comparing the above stress tensor $\bT_{ab}$ with the general stress tensor $T^{_{(R)}}_{ab}$ in \eqref{TRab}, one can read out exactly the same coefficients in \eqref{coefficientsGB} when $\a=0$.
Thus, through using the Hamiltonian constraint and Petrov type \I condition, we recover the Brown-York stress tensor \eqref{TabGB} dual to the bulk metric in (\ref{metricGB}) in the case of Einstein gravity.

\subsection{Recover the Rindler fluid in Einstein-Gauss-Bonnet gravity}
\label{SecPetrovGB}

In this subsection, we will recover the Rindler fluid in Einstein-Gauss-Bonnet gravity.
For the convenience of calculation and since $\bHa\equiv0$, we write the Hamiltonian constraint \eqref{ttHaGB} as
\begin{align}
\ttHaGB&=\tHa^{(\a)}+\de\bHa^{\upH}=\de\tHa^{(\a)}+\de\bHa^{\upH}=0,\label{HaGB2}
\end{align}
where $\tHa^{(\a)}$ and $\de\bHa^{\upH}$ can be found in \eqref{Hama} and \eqref{tHaH}, respectively.
Since $\bPe_{ab}\equiv 0$, the Petrov type \I condition in \eqref{ttPeGB} becomes
\begin{align}
\ttPeGB_{ab} &=\tPe^{(\a)}_{ab}+\de\bPe^{_{(H)}}_{ab}=\de\tPe^{(\a)}_{ab}+\de\bPe^{_{(H)}}_{ab}=0,
\label{PabGB2}
\end{align}
where $\tPe^{(\a)}_{ab}$ and $\de\bPe^{_{(H)}}_{ab}$ can be found in \eqref{Paba} and \eqref{tPabH}.
On the other hand, from \eqref{KabTab} and with the results in \eqref{bTab2}, one has
\begin{align}\label{bKab}
2\bK_{ab}&=-(\pp+\erho^{(2)}) u_{a}u_{b}-\Pi^{(1)}_{ab} -\Pi^{(2)}_{ab}+O(\p^3).
\end{align}
We then assume the following decomposition of the extrinsic curvature
\begin{align}
K_{ab}&=\,\vr\, u_a u_b + \vp_{ab},\qquad \qquad~~\vr\equiv K_{ab}u^a u^b,\qquad\quad~~\vp_{ab}\equiv h_a^c h_b^d K_{cd},\label{Kabrhopi}\\
\de K^{(\a)}_{ab}&=\dvr^{(\a)} u_a u_b + \dvp^{(\a)}_{ab},\quad \dvr^{(\a)}\equiv \dK^{(\a)}_{ab}u^a u^b,\quad \quad\dvp^{(\a)}_{ab}\equiv h_a^c h_b^d \dK^{(\a)}_{cd}.\label{dKabrhopi}
\end{align}
 From \eqref{TKbK}, we then conclude
\begin{align}\label{rhopi}
2\vr&=-\pp-\erho^{(2)} +2\dvr^{(\a)}+O(\p^3)+O(\a^2),\quad \\
2\vp_{ab}&= -\Pi^{(1)}_{ab} -\Pi^{(2)}_{ab} +2\dvp^{(\a)}_{ab}+O(\p^3)+O(\a^2).\label{rhopi1}
\end{align}
Putting \eqref{bKab} into \eqref{tHaH} and \eqref{Paba}, one has
\begin{align}\label{deHaH}
\de\bHa^{_{^{^{{(H)}}}}}=O(\p^3),\quad
\de\bPe^{_{(H)}}_{ab}=-6\a\pp^2\[\Oo_{ac}\Oo^c_{~b}+ p^{-1} h_{a b}\Oo_{cd}\Oo^{cd}\]+O(\p^3).
\end{align}
As  the Gauss-Bonnet corrections to Hamiltonian constraint and Petrov type \I condition appear at the second order in the derivative expansion, we only need to consider the second order corrections
that $\dvr^{(\a)}\sim\dvp^{(\a)}_{ab}\sim O(\p^2)$. Thus put \eqref{Kabrhopi} into \eqref{Hama} and \eqref{Paba}, we have
\begin{align}
\Ha^{(\a)}&=(2\vr-\vp)\vp +\vp_{ab}\vp^{ab},\label{HaIa}\\
\tPe^{(\a)}_{ab}&=(\vp-2\vr)\vp_{ab}-\vp_{ac}\vp^c_{~b}+2\vr\K_{ab}+2\K_{(a}^{~c}\vp_{b)c}+2\Oo_{(a}^{~c}\vp_{b)c}+2h_{a}^c h_b^d D \vp_{cd}.\label{PabIa}
\end{align}
Taking into account of \eqref{rhopi} and \eqref{rhopi1} and consider the first order in the small $\a$ expansion, we obtain
\begin{align}\label{deHaa}
\de\tHa^{(\a)}=\Ha^{(\a)}=-\pp\de\pi^{(\a)},\quad
\de\tPe^{(\a)}_{ab}=\tPe^{(\a)}_{ab}=\pp\de\pi^{(\a)}_{ab}.
\end{align}
With \eqref{deHaH} and \eqref{deHaa},  at the second order in the derivative expansion, the  Hamiltonian constraint leads to
\begin{align}
\tHa^{(2)}=\de\tHa^{(\a)}+\de\bHa^{\upH}=&0\,\Rightarrow\, \de\pi^{(\a)}=0.
\end{align}
And the Petrov type I condition leads to
\begin{align}
\tPe_{ab}^{(2)}=\de\tPe_{ab}^{(\a)}+\de\bPe_{ab}^{_{(H)}}=&0\Rightarrow \dvp^{(\a)}_{ab}=6\a\pp \[\Oo_{ac}\Oo^c_{~b}+ p^{-1} h_{a b}\Oo_{cd}\Oo^{cd}\].
\end{align}
We can see that there is no constraint on $\vr^{(\a)}$ at this order, and it will be determined by the gauge choice of the stress tensor.
Then from \eqref{dTabJ}, we obtain
\begin{align}
\de\tT^{(\a)}_{ab}=&-2\dvp^{(\a)}u_au_b+2(\dvp^{(\a)}-\dvr^{(\a)})h_{ab}-2\dvp^{(\a)}_{ab}.
\end{align}
On the other hand, a straightforward  calculation from \eqref{dTabJ} and \eqref{bKab} gives
\begin{align}
\de\bT^{_{(J)}}_{ab}=&\a\pp\[-\Pi^{(1)}_{ac}\Pi^{c(1)}_{~b}+\frac{1}{2}\(\Pi^{(1)}_{cd}\Pi_{(1)}^{cd}\)h_{ab}\],
\end{align}
where $\Pi^{(1)}_{ab}$ has been obtained in \eqref{Piab1}.
Put them together, we obtain
\begin{align}\label{deTab}
\dT_{ab}=\de\tT^{(\a)}_{ab}+\de\bT^{_{(J)}}_{ab}= & -4\a\pp\(\K_{ac}\K^c_{~b}+3 p^{-1}\Oo_{ac}\Oo^c_{~b}\)\nn\\
&+\[-2\de\vr^{(2)}+2\a\pp\(\K_{cd}\K^{cd}-6p^{-1}\Oo_{cd}\Oo^{cd}\)\]h_{ab}.
\end{align}
The isotropic gauge of the pressure leads to $\de\vr^{(2)}=\a\pp\(\K_{cd}\K^{cd}-6p^{-1}\Oo_{cd}\Oo^{cd}\)$.
Then the stress tensor from Petrov type \I condition turns out to be $\bT_{ab}+\dT_{ab}$ with \eqref{Tcovab2} and \eqref{deTab},
which match exactly with the $\ttTGB_{ab}$ in \eqref{TabGB} from the fluid/gravity calculation.

\section{The non-relativistic hydrodynamic expansion}
\label{Nonrelativistic}

The Rindler fluid with Gauss-Bonnet corrections in the following non-relativistic hydrodynamic expansion has been studied in \cite{Compere:2011dx,Chirco:2011ex}
\begin{align}\label{orders}
v_i\sim \e,~~~~P \sim \e^2,~~~~\p_i \sim \e,~~~~\p_\tau \sim \e^2.
\end{align}
And the dual tress tensor turns out to be $\ttT_{ab}=\bT_{ab}+\de\tT_{ab}$, where $\bT_{ab}$ come from the Einstein sector, which are given by \cite{Compere:2011dx},
\begin{align}
{\bT^\tau}_{~i} =& +r_c^{-3/2}v_i+r_c^{-5/2}\[v_i(v^2+P)-2 r_c\s_{ij}v^j\]+O(\eps^5),\nn\\
{\bT^\tau}_{~\tau} =& -r_c^{-3/2}v^2 -r_c^{-5/2}\left[v^2(v^2+P)-2r_c \s_{ij}v^iv^j-{2 r_c^2}\s_{ij}\s^{ij}\right]+O(\eps^6),\nn\\
{\bT}_{\,ij} =& + r_c^{-1/2}\,\delta_{ij}+ r_c^{-3/2}\[P\delta_{ij}+v_iv_j-2  r_c \s_{ij}\]\nn\\
   &+ r_c^{-5/2}\[ v_iv_j (v^2+P)-{  r_c}\s_{ij}v^2+2 r_c v_{(i}\p_{j)}P -  r_c v_{(i}\p_{j)}v^2-2 r_c^2 v_{(i}\p^2 v_{j)} \right.\nn\\[1ex]
   &-2r_c^2\s_{ik}{\s^k}_{j}-4 r_c^2{\s}_{k(i}\omega^k_{~j)}-4 r_c^2 \omega_{ik}{\omega^k}_{j}
   -4 r_c^2 \p_i\p_j P  +3 r_c^3 \p^2 \s_{ij}]+O(\eps^6), \nn \\
 \bT =& \, \,{\bT^\tau}_{~\tau}+{\bT^i}_{~i}= p r_c^{-1/2}+p r_c^{-3/2}P+O(\eps^6).\label{Tt}
\end{align}
Here the fluid shear $\sigma_{ij}=\p_{(i}v_{j)}$ and vorticity $\omega_{ij}=\p_{[i}v_{j]}$.
And $\de\tT_{ab}$ come from the Gauss-Bonnet term, with the non-vanishing components \cite{Chirco:2011ex,Eling:2012xa},
\begin{align}\label{dehTij}
\de\tT_{ij}&=-4\a r^{-3/2}_c \(\s_{ik}{\s^k}_{j} +3 \,\omega_{ik}{\omega^k}_{j}\)+O(\eps^6),\\
\de\tT&=\de^{ij}\de\tT_{ij}
       =-4\a r^{-3/2}_c \(\s_{ij}\s^{ij} -3 \,\omega_{ij}\omega^{ij}\)+O(\eps^6).\label{dehTt}
\end{align}
We can see that the contributions from the Gauss-Bonnet term only appear at order $\e^4$. This comes from the fact that the first non-zero components of the Riemann tensor appear at order $\epsilon^2$~\cite{Chirco:2011ex}. And notice that the situation for the case of Einstein gravity has been studied in \cite{Cai:2013uye}.
Thus  we  need only to focus on the Gauss-Bonnet corrections to the Petrov type \I condition and Hamiltonian constraint at $\ep^4$ in this section.

\subsection{Petrov type \I condition in Rindler fluid}

Introduce the new coordinate $x^0 =\sqrt{r_c } \tau$,
the flat induced metric $\gamma_{ab}$ in \eqref{RindlerCut} becomes
\begin{align}
\d s^2_{p+1} &=\eta_{ab}\d x_a\d x^b= -(\d{x^0})^2+\delta_{ij}\d x^i\d x^j.
\end{align}
The $(p+2)$ Newman-Penrose-like vector fields are given with respect to the ingoing and outgoing pair of null vectors as \cite{Lysov:2011xx}
\begin{align}\label{frame1}
\sqrt{2}\ell = \p_0 -n,\quad\sqrt{2}k=-\p_0-n,\quad m_i=\p_i.
\end{align}
Here $n$ is the unit normal vector of the hypersurface \sc, $\p_0$ and $\p_i$ are the tangent vectors to \sc. The spacetime is at least Petrov type \I if
\begin{align}\label{petrovij}
\Pij_{ij}\equiv 2C_{(\ell)i(\ell)j}=0,\quad C_{(\ell)i(\ell)j}\equiv \ell^\mu m_i^\nu \ell^\alpha m_j^\beta C_{\mu\nu\alpha\beta}.
\end{align}
With the Guass-Codazzi equations given in \eqref{typeI}, we have the Petrov type \I condition up to the first order in the small $\a$ expansion as
\begin{align}\label{typeIij}
\PijGB_{ij}&=\bPij_{ij}+\de\Pij^{(\a)}_{ij}+\de\bPij^{_{^{(H)}}}_{ij}=0,\\
\bPij_{ij} &\equiv-\bM^{\pe}_{ij}+2\bN^{\pe}_{0ij}+\bM^{\pe}_{0i0j},\quad
\de\Pij^{(\a)}_{ij} \equiv-\de\M^{\pe}_{ij}+2\de\N^{\pe}_{0ij}+\de\M^{\pe}_{0i0j},\label{dPija}
\end{align}
with
\begin{align}
\de\bPij^{_{(H)}}_{ij}=&- 2\a \bH^{\bot}_{ij} + 2\a p^{-1}\delta_{ij}\[\bH_{\mu\nu} n^\mu n^\nu-2\bH_{0 \mu}n^\mu+\bH_{ 0  0 }+\bH\]\\
=&-2\a \(\bM_{i}^{~cde}\bM_{j cde}+2\bN_{i}^{~cd}\bN_{j cd}+\bN^{cd}_{~~i}\bN_{cd\,j}+2\bM_{i}^{~d}\bM_{j d}\) \nn\\
&+\a p^{-1} \de_{ij}\Big[2\(\bM_{ 0 }^{~cde}\bM_{ 0  cde}+2\bN_{ 0 }^{~cd}\bN_{ 0  cd}+\bN^{cd}_{~~ 0 }\bN_{cd  0 }+2\bM_{ 0 }^{~d}\bM_{ 0  d}\)\nn\\
& +4(\bM_{ 0 cde}\bN^{dec}- 2 \bM^{cd} \bN_{ 0 cd}) +\(\bM^{cdef}\bM_{cdef}+6\bN^{cde} \bN_{cde} +8\bM^{cd}\bM_{cd}\)\Big].\label{dPijH}
\end{align}
The Hamiltonian constraint becomes
\begin{align}\label{Hcut}
\HijGB&=\bHij+\de\Hij^{(\a)}+\de\bHij^{\upH}=0,\\
\bHij&\equiv\bM,\qquad \de\Hij^{(\a)}\equiv\de\M,\label{dHija}
\end{align}
with
\begin{align}
\de\bHij^{\upH}\equiv- 4 \a \bH_{\mu\nu}n^{\mu}n^{\nu}=
 \a\(-4\bM_{ab}\bM^{ab}+\bM_{abcd}\bM^{abcd}\).\label{dHijH}
\end{align}

Notice that the frame choice in \eqref{frame1} singles out a preferred time coordinate $\p_0$ and thus breaks Lorentz invariance. It has been shown in \cite{Cai:2013uye} that  with the frame \eqref{frame1}, the Petrov type \I condition for vacuum Einstein gravity $\bPij_{ij}=0$ is violated at order $\ep^4$:
\begin{align}\label{bPijE}
\bPij^{_{(E)}}_{ij}=\bPij_{ij}=\frac{1}{2}r_c^{-3}\[6 r_c v_k v_{(i}\o^k_{~j)}-2r_c^2 v_{(i}\p^2v_{j)}-4 r_c^2 v^k\p_{(i}\o^k_{~j)}+r_c^3\p^2\sigma_{ij}\]+O(\ep^6).
\end{align}
However, after straightforward calculations with the stress tensor \eqref{Tt} and \eqref{dehTij}, we find
\begin{align}
\de\bHij^{\upH}&=\de\Hij^{(\a)}=O(\ep^6),\label{deHH}\\
\de\bPij^{_{(H)}}_{ij}&=-\de\Pij^{(\a)}_{ij}=-6\a r_c^{-2}\(\omega_{ik}\omega^k_{~j}+ p^{-1} \de_{ij}\omega_{kl}\omega^{kl}\)+O(\ep^5).\label{dePijH}
\end{align}
Thus, there are no Gauss-Bonnet corrections to the Hamiltonian constraint \eqref{Hcut} and Petrov type \I condition \eqref{typeIij} up to order $\ep^4$ and up to the first order in small $\a$. In the following subsection, we will show that either demand $\bPij_{ij}=0$ or with the stress tensor  (\ref{Tt}) of Rindler fluid in vacuum Einstein gravity, and impose
\begin{align}\label{PHij}
\de \Hij=\de\Hij^{(\a)}+\de\bHij^{\upH}=0,\quad
~\de \Pij_{ij}=\de\Pij^{(\a)}_{ij}+\de\bPij^{_{(H)}}_{ij}=0,
\end{align}
we can get exactly the contribution \eqref{dehTij} of the Gauss-Bonnet term to the stress tensor of the dual fluid, without solving the Einstein-Gauss-Bonnet field equations.

\subsection{Recover the Gauss-Bonnet corrections}

If we still demand the Petrov type \I condition $\bPij_{ij}=0$ in the vacuum Einstein gravity, it has been shown in \cite{Cai:2013uye} that the stress tensor in \eqref{Tt} can be recovered up to an additional term at $\ep^4$:
\begin{align}
\de\bT^{_{(E)}}_{ij}
=r_c^{-5/2}\[6 r_c v_k v_{(i}\o^k_{~j)}-2r_c^2 v_{(i}\p^2v_{j)}-4 r_c^2 v^k\p_{(i}\o^k_{~j)}+r_c^3\p^2\sigma_{ij}\]+O(\ep^6).
\end{align}
Then using $\bT_{ab}+\de\bT^{_{(E)}}_{ab}$ instead of $\bT_{ab}$ in \eqref{Tt},
we can obtain the  extrinsic curvature $\bK_{ab}$ from \eqref{KabTab},
and then put them into \eqref{dHijH} and \eqref{dPijH},
which lead to the same results in \eqref{deHH} and \eqref{dePijH}, we see that
\begin{align}
\de\bHij^{\upH}&=O(\ep^6),\label{HHij2}\\
\de\bPij^{_{(H)}}_{ij}&=-6\a r_c^{-2}\(\omega_{ik}\omega^k_{~j}+ p^{-1} \de_{ij}\omega_{kl}\omega^{kl}\)+O(\ep^5).\label{PHij2}
\end{align}
They are not affected by the additional term $\de\bT^{_{(E)}}_{ab}$.
To cancel the non-vanishing $\de\bPij^{_{(H)}}_{ij}$ at order $\ep^4$ in \eqref{PHij2}, we assume $\de\tT_{ab}\sim O(\ep^4)$ such that $\de\Hij^{(\a)}$ in \eqref{dHija} and  $\de\Pij_{ij}^{(\a)}$ in \eqref{dPija} also appear at order $\ep^4$. As  ${\bT^{\tau}}_{~i}$ in \eqref{Tt} has been fixed through the  frame choice  of the velocity~\cite{Cai:2013uye}, we only need to set the Gauss-Bonnet correction $\de{\tT^{\tau}}_{i}=O(\ep^5)$. Then put the relation \eqref{dKabTab} into \eqref{dHija} and \eqref{dPija}, we obtain
\begin{align}
\de \Hij^{(\a)}&=\frac{1}{2} r_c^{-1/2}\[-{\de T^\tau}_\tau+4\a\(\bJ-3{\bJ^\tau}_{~\tau}\)\],\label{dTtt}\\
\de \Pij_{ij}^{(\a)}&=\frac{1}{2} r_c^{-1/2}\[-{\de T}_{\,ij}-4\a\(3  \bJ_{ij}-2 p^{-1}\bJ \de_{ij}\)+p^{-1}\de T \de_{ij}\].\label{dTij}
\end{align}
With \eqref{Jab1},\eqref{KabTab} and \eqref{Tt}, we have the non-zero components of $\bJ_{ab}$ as
\begin{align}\label{bJij}
{\bJ^\tau}_{~\tau}=\frac{1}{6}\,r_c^{-3/2}(\sigma_{ij}\sigma^{ij})+O(\ep^6),\quad
{\bJ}_{ij}=\frac{1}{3}\,r_c^{-3/2}\sigma_{ik}\sigma^{k}_{~j}+O(\ep^6),\quad
{\bJ}={\bJ^\tau}_{~\tau}+{\bJ^i}_{~i}\,.
\end{align}
Substituting them into \eqref{PHij}, we finally obtain
\begin{align}
{\de T^{\tau}}_\tau&=O(\ep^6),\label{Tttresult}\\
{\de T}_{\,ij}&=-4\a r^{-3/2}_c \(\s_{ik}{\s^k}_{j} +3 \,\omega_{ik}{\omega^k}_{j}\)\nn\\
&\quad+p^{-1}\[\de T+4\a r^{-3/2}_c \(\s_{ij}{\s}^{ij} -3 \,\omega_{ij}{\omega}^{ij}\)\] \de_{ij}+O(\ep^6).\label{Tijresult}
\end{align}
After choosing the isotropic gauge such that there are no corrections to the $\de_{ij}$ part of the stress tensor at this order as in \cite{Compere:2011dx,Chirco:2011ex},
we have $\de T=-4\a r^{-3/2}_c \(\s_{ij}{\s}^{ij} -3 \,\omega_{ij}{\omega}^{ij}\)$.
These results exactly match with the Gauss-Bonnet corrections in the stress tensor of Rindler fluid which are given in \eqref{dehTij} and \eqref{dehTt} from the fluid/gravity calculation.

 Alternatively, once the stress tensor $\bT_{ab}$ of Rindler fluid is given in vacuum Einstein gravity (\ref{Tt}) from the fluid/gravity calculation, by demanding the condition \eqref{PHij} to hold  that the additional Gauss-Bonnet corrections to the Hamiltonian constraint and Petrov type I condition vanish,
one can show that the formulas between \eqref{HHij2} and \eqref{Tijresult} are the same as the those in the case by using $\bT_{ab}+\de\bT^{_{(E)}}_{ab}$, such that we can again obtain the Gauss-Bonnet corrections to the stress tensor of the Rindler fluid in \eqref{dehTij} for the Einstein-Gauss-Bonnet gravity.

\section{Conclusion}
\label{Conclusion}

To summarize, we have checked the Petrov type I condition for the vacuum solutions of Einstein-Gauss-Bonnet gravity in both relativistic and non-relativistic hydrodynamic expansions.
With the solution constructed in \cite{Eling:2012xa}, we have shown that the spacetime  is at least Petrov type \I up to the second order in the relativistic hydrodynamic expansion. 
Turn the logic around, assuming the Hamiltonian constraint and Petrov type \I condition on a
finite cutoff hypersurface, we have shown that the dual stress tensor can be recovered with correct first and second order transport coefficients by taking the Gauss-Bonnet coefficient as an expansion parameter.
While in the non-relativistic hydrodynamic expansion~\cite{Chirco:2011ex}, although the Petrov type I condition is violated at order $\ep^4$ in the vacuum Einstein gravity \cite{Cai:2013uye}, we have found that the Gauss-Bonnet term does not contribute to the violation terms in the Petrov type I condition up to $\ep^4$.
Thus, given the stress tensor of the Rindler fluid in vacuum Einstein gravity, we have shown that demanding the additional Gauss-Bonnet corrections to the Petrov type I condition and Hamiltonian constraint vanish at the first order of $\a$ expansion, the Gauss-Bonnet corrections to the stress tensor of dual fluid can also be recovered.

Notice that in both cases, in order to recover the stress tensor of dual fluid from the Petrov type I condition, we have additionally taken the small $\a$ limit. And up to the first order of $\a$ expansion, the Petrov type I condition can be expressed as a function of extrinsic curvature and other intrinsic quantities on the hypersurface. Actually, note the fact that the Einstein-Gauss-Bonnet field equations are quasi-linear in terms of $\a$~\cite{Deruelle:2003ck,Maeda:2003vq}, and the dual stress tensor with Gauss-Bonnet corrections in \eqref{TabGB} is also linear in terms of $\a$. It is not surprised that we can still recover the stress tensor \eqref{TabGB} even when we take the small $\a$ limit.

So far most of studies on the Petrov type \I condition has been focused on the case with asymptotically flat spacetimes. It is quite important and interesting to investigate corresponding ones for asymptotically AdS spacetimes based on the AdS/CFT correspondence,
as the regularity condition on the future horizon of spacetime is necessary and important for the perturbations in the fluid/gravity correspondence and imposing the Petrov type I condition on the spacetime is mathematically much simpler than directly solving the
perturbative gravitational field equations in order to find the stress tensor of dual fluid.
 On the other hand, the KSS bound \cite{Kovtun:2003wp} states that the universal value of the ratio of shear viscosity over entropy density from the AdS/CFT calculation is always above $\eta/s=1/4\pi$, while in the AdS gravity with curvature squared corrections, the bound is found to be violated by the Gauss-Bonnet term~\cite{Kats:2007mq,Brigante:2007nu,Brigante:2008gz}.
With the static black brane solution in \cite{Cai:2001dz},  it is expected that the universal value with Gauss-Bonnet correction $\eta/s=[1-2(p+1)(p-2)\a]/4\pi$ can also be recovered from the Petrov type I condition on the dual fluid.

\section*{Acknowledgments}
This work is supported by National Natural Science Foundation of China
(No.10821504, No.11035008, and No.11375247).
We thank L.~Li for helpful conversation.
Y. L.~Zhang thanks Professors K. Skenderis and M. Taylor for many valuable discussions on this topic, as well as C. Eling and A. Meyer for helpful correspondence.
\appendix

\allowdisplaybreaks

\section{Classification of the Weyl tensor}
\label{APetrovtype}
In four dimensional spacetime, tensor classification plays an important role in studying the exact solutions of Einstein field equations~\cite{Stephani:2003tm}. And in particular, the Petrov type classification of Weyl tensor has interesting physical applications. It has been generalized to the arbitrarily higher dimensional spacetimes in~\cite{Milson:2004jx}. In this appendix, we briefly summarize these results based on~\cite{Coley:2004jv,Coley:2007tp},  which can also be reduced to the Petrov type classification in four dimensions.

Consider a $p+2$ dimensional Lorentz manifold ($p\ge2$) with signature $(- + ... +)$ and choose a null frame $\hell,\, \hk, \, \hm_i$, which satisfies the following orthogonal and normalization conditions
\begin{align}
\hell^2=\hk^2=0,~~(\hk,\hell)=1,~~(\hm_i,\hk)=(\hm_i,\hell)=0,~~(\hm_i,\hm_j)=\delta_{ij},
\end{align}
so that in this frame the metric of the manifold can be decomposed as
\begin{align}
g_{\mu\nu}=2 \hell_{(\mu} \hk_{\nu)} +\de_{ij}\hm^i_{\;\mu}\hm^j_{\;\nu},\qquad
g^{\mu\nu}=2 \hell^{(\mu} \hk^{\nu)} +\de^{ij}\hm_i^{\;\mu}\hm_j^{\;\nu}.
\end{align}

The null frame is covariant under the following boost transformation,
\begin{align}
  \label{eq:boost}
    \hell &\rightarrow \lambda\,\hell,\quad {\hk}\rightarrow \lambda^{-1} \hk,\quad
    \hm_i\rightarrow \hm_i,     \quad \lambda \neq 0.
\end{align}
For a rank $q$ tensor $T$ on the manifold, its components $T_{\mu_1... \mu_q}$ with fixed list of indices are null frame scalars, and they transform under the boost transformation as
\begin{equation}
T_{\mu_1... \mu_q}\rightarrow\lambda^{b_{\{\mu\}}} T_{\mu_1... \mu_q},
\quad b_{\{\mu\}}=b_{\mu_1}+...+b_{\mu_q},~ b_{(\bmell)}=1,~b_i=0,~b_{(\bmk)}=-1.
\end{equation}
$b$ is named as the boost-weight of the null-frame scalar $T_{\mu_1... \mu_q}$.
The boost order (along $\hell$) of the tensor $T$ is defined to be the largest value of $b_{\{\mu\}}$ among all the non-vanishing components $T_{\mu_1... \mu_q}$. It is only a function of the null direction $\hell$ and is denoted as $\mbhl$.

The Weyl tensor can be decomposed and sorted by the boost weight of its components,
\begin{align}\label{weyl}
 C_{\a\b\g\de} &=C^{[2]}_{\a\b\g\de}+C^{[1]}_{\a\b\g\de}+C^{[0]}_{\a\b\g\de}+C^{[-1]}_{\a\b\g\de}+C^{[-2]}_{\a\b\g\de},
 \end{align}
where the superscript index indicates the boost weight and
\begin{align}
C^{[2]}_{\a\b\g\de}&=4C_{(\bmell)i(\bmell)j} \hk_{\{\a}{\hm^i}_\b \hk_\g {\hm^j}_{\de\}},\nn\\
C^{[1]}_{\a\b\g\de}&= 8C_{(\bmell)(\bmk)(\bmell)i} \hk_{\{\a} \hell_\b \hk_\g {\hm^i}_{\de\}}+ 4C_{(\bmell)ijk} \hk_{\{\a} {\hm^i}_\b {\hm^j}_\g {\hm^k}_{\de\}}, \nn \\
C^{[0]}_{\a\b\g\de}&=4C_{(\bmell)(\bmk)(\bmell)(\bmk)} \hk_{\{\a} \hell_\b \hk_\g \hell_{\de\}} + 4C_{(\bmell)(\bmk)ij}
 \hk_{\{\a} \hell_\b {\hm^i}_\g {\hm^j}_{\de\}}\nn\\
 &+8C_{(\bmell)i(\bmk)j} \hk_{\{\a} {\hm^i}_\b \hell_\g {\hm^j}_{\de\}} +
 C_{ijkl} \hm^i_{\{\a} {\hm^j}_\b  {\hm^k}_\g {\hm^l}_{\de\}},\nn\\
C^{[-1]}_{\a\b\g\de}&=8C_{(\bmk)(\bmell)(\bmk)i} \hell_{\{\a} {\hk}_\b \hell_\g {\hm^i}_{\de\}} + 4C_{(\bmk){ijk}} \hell_{\{\a} {\hm^i}_\b {\hm^j}_\g {\hm^k}_{\de\}},\nn\\
C^{[-2]}_{\a\b\g\de}&=4C_{(\bmk)i(\bmk)j} \hell_{\{\a} {\hm^i}_\b \hell_\g {\hm^j}_{\de\}}.
 \end{align}
The notations $T_{\{\a\b\g\de\}}\equiv(T_{[\a\b][\g\de]}+T_{[\g\de][\a\b]})/2$, as well as $C_{(\bmell)i(\bmk)j}\equiv C_{(\bmell)i(\bmell)j}\hell^\mu{\hm_j}^\a \hk^\nu{\hm_j}^\b$ and so on, have been introduced. The Weyl tensor is generically of boost order $\mbhl=2$, and a null vector $\hell$ is defined to be aligned with the Weyl tensor whenever $\mbhl\le 1$. In this case, $\hell$ is a Weyl aligned null direction, and $1-\mbhl\in\{0,1,2,3\}$ is the order of alignment. It usually depends on the rank and symmetry properties of the tensors.

According to \cite{Milson:2004jx}, the principal type of the Weyl tensor in a Lorentzian manifold is I, II,
III, N according to whether there exists an aligned $\hell$ of alignment order $0,1,2,3$, respectively. If no aligned $\hell$ exists, the manifold is of (general) type G, if the Weyl tensor vanishes the manifold is of type O. The
algebraically special types with necessary condition are summarized as follows:
\begin{align}
&  \text{Type I}: ~~~C_{(\bmell)i(\bmell)j}=0, \nn\\
&  \text{Type II}: ~~C_{(\bmell)i(\bmell)j}=C_{(\bmell)ijk}=0, \nn\\
&  \text{Type  III}:~C_{(\bmell)i(\bmell)j}=C_{(\bmell)ijk}=C_{ijkl} =C_{(\bmell)(\bmk)ij}=0, \nn\\
&  \text{Type N}:  ~~C_{(\bmell)i(\bmell)j}=C_{(\bmell)ijk}=C_{ijkl} = C_{(\bmell)(\bmk)ij}=C_{(\bmk)ijk}=0.
\end{align}
Following the curvature tensor symmetries and the trace-free condition \cite{Coley:2004jv},
one can reach some familiar Petrov types with the following properties,
\begin{align}
&\text{Type I}: ~~~ C^{[2]}_{\a\b\g\de}=0,
\nn\\&  \text{Type II}: ~~C^{[2]}_{\a\b\g\de}=C^{[1]}_{\a\b\g\de}=0,
\nn\\&  \text{Type D}:  ~~C^{[2]}_{\a\b\g\de}=C^{[1]}_{\a\b\g\de} =C^{[-1]}_{\a\b\g\de}=C^{[-2]}_{\a\b\g\de}=0,
\nn\\&  \text{Type  III}:~C^{[2]}_{\a\b\g\de}=C^{[1]}_{\a\b\g\de} =C^{[0]}_{\a\b\g\de}=0,
\nn\\&  \text{Type N}:  ~~C^{[2]}_{\a\b\g\de}=C^{[1]}_{\a\b\g\de} =C^{[0]}_{\a\b\g\de}=C^{[-1]}_{\a\b\g\de}=0,
\nn\\&  \text{Type O}: ~~C^{[2]}_{\a\b\g\de}=C^{[1]}_{\a\b\g\de} =C^{[0]}_{\a\b\g\de}=C^{[-1]}_{\a\b\g\de}=C^{[-2]}_{\a\b\g\de}=0.
\end{align}
Further  classifications in more detail can be found in~\cite{Coley:2004jv,Coley:2007tp}.

\section{Decomposition of the Riemann tensor}
\label{Projections}

The Riemann tensor and its contractions can be decomposed along and perpendicular to
a spacelike unit normal vector $n$,
\begin{align}\label{Rabcd}
g_\m^\a g_\n^\b g_\s^\g g_\l^\de R_{\a\b\g\de}
&=\,\M_{\m\n\s\l}-n_\m \N_{\s\l\n} +n_\n \N_{\s\l\m} -n_\s N_{\m\n\l} +n_\l \N_{\m\n\s} \nn\\
&\quad\,+n_\m n_\s \L_{ \n \l}-n_\m n_\l \L_{ \n\s }+ n_\n n_\l \L_{\m \s }-n_\n n_\s\L_{\m \l},\nn\\
g_\m^\a g_\n^\b R_{\a\b}&=\M_{\m\n}+n_\m \N_\n+n_\n \N_\m+\L_{\m\n}+ n_\m n_\n\L,\nn\\
R&=\,\M+2\L=-\M+2\g^{\b\de} R_{\b\de},
\end{align}
where we have defined the following notations with $\g_{\mu\nu}=g_{\mu\nu}-n_\mu n_\nu$,
\begin{align}
\M_{\m\n\s\l}&\equiv\g_\m^\a \g_\n^\b\g_\s^\g \g_\l^\delta R_{\a\b\g\de},\quad
\N_{\m\n\s}\equiv\g_\m^\a \g_\n^\b \g_\s^\g n^\de R_{\a\b\g\de},\quad
\L_{\m\n}\equiv\g_\m^\a n^\b \g_\n^\g n^\de R_{\a\b\g\de},\nn\\
\M_{\m\n}&\equiv\g^{\a\b} \M_{\m\a\n\b},\qquad
M\equiv\g^{\a\b}\M_{\a\b},\qquad
\N_{\m}\equiv \g^{\a\b}\N_{\a\m\b},\qquad
\L\equiv \g^{\a\b}\L_{\a\b}.
\end{align}
One can also obtain the decomposition of their combinations, such as,
\begin{align}
R_{\m}^{~\s\l\r }R_{\n\s\l\r} n^\m n^\n&=N^{cde} N_{cde} +2\L^{cd}\L_{cd}, \nn\\
R_{\m}^{~\s\l\r }R_{\n\s\l\r} n^\m h^\n_b&=-M_{b cde}N^{dec} -2 \L^{cd}N_{b cd},\nn\\
R_{\m}^{~\s\l\r }R_{\n\s\l\r} h^\m_a h^\n_b&=M_{a}^{~cde}M_{b cde}+2N_{a}^{~cd}N_{b cd}+N^{cd}_{~~a}N_{cd b}+  2 \L_{a}^{~c} \L_{c b},\nn\\
R_{\m}^{~\s\l\r }R_{\n\s\l\r}g^{\m\n}&=M^{cdef}M_{cdef}+4N^{cde} N_{cde} +4\L^{cd}\L_{cd}.
\end{align}
Then $\bH_{\m\n}\equiv\bR_{\mu}^{~\sigma\lambda\rho}\bR_{\nu \sigma\lambda\rho}-\frac{1}{4}\(\bR^{\kappa\sigma\lambda\rho}\bR_{\kappa \sigma\lambda\rho}\)\bg_{\m\n}$
in \eqref{bHmn} can be decomposed as
\begin{align}\label{bHuu}
\bH_{(\iu n\iu)(\iu n\iu)}&\equiv \bH_{\m\n} n^\m n^\n
=\bL^{cd}\bL_{cd}-\frac{1}{4}\bM^{cdef} \bM_{cdef}, \nn\\
\bH_{(\iu n\iu)(\iu u\iu)}&\equiv \bH_{\m\n} n^\m \g^\n_b u^b
=-\bM_{(\iu u\iu)cde}\bN^{dec} -2 \bL^{cd} \bN_{(\iu u\iu)cd},\nn\\
\bH_{(\iu u\iu)(\iu u\iu)}&\equiv \bH_{\m\n}  \g^\m_a \g^\n_b u^a u^b
=\bM_{(\iu u\iu)}^{~~cde}\bM_{(\iu u\iu) cde}+2\bN_{(\iu u\iu)}^{~cd}\bN_{(\iu u\iu) cd}+\bN^{cd}_{~~(\iu u\iu)}\bN_{cd (\iu u\iu)}+2\bL_{(\iu u\iu)}^{~~d}\bL_{(\iu u\iu) d}\nn\\
&\qquad\qquad\qquad\quad\quad+\frac{1}{4}\(\bM^{cdef}\bM_{cdef}+4\bN^{cde} \bN_{cde} +4\bL^{cd}\bL_{cd}\),\nn\\
\bH^{\pe}_{ab}&\equiv \bH_{\m\n}  \g^\m_c \g^\n_d h_a^c h_b^d
= h_a^mh_b^n\(\bM_{m}^{~cde}\bM_{n cde}+2\bN_{m}^{~cd}\bN_{n cd}+\bN^{cd}_{~~m}\bN_{cd n}+2\bL_{m}^{~d}\bL_{n d}\)\nn\\
&\qquad\qquad\qquad\quad\quad-\frac{1}{4}\(\bM^{cdef}\bM_{cdef}+4\bN^{cde} \bN_{cde} +4\bL^{cd}\bL_{cd}\) h_{a b},\nn\\
\bH&\equiv \bH_{\m\n}  g^{\m\n}
=-\frac{p-2}{4}\(\bM^{cdef}\bM_{cdef}+4\bN^{cde} \bN_{cde} +4\bL^{cd}\bL_{cd}\).
\end{align}

\end{document}